\documentstyle[aaspp4]{article}

\begin{document}

\title{Dwarf galaxy formation induced by galaxy interactions}

\author{Tadashi Okazaki,  \& Yoshiaki Taniguchi}

\affil{Astronomical Institute, Graduate School of Science,
       Tohoku University, Aramaki, Aoba, Sendai 980-8578, Japan}

\begin{abstract}
Growing evidence has been accumulated for that
some gas-rich dwarf galaxies are formed 
from material liberated by galaxy collisions and/or mergers.
Also, gas-poor dwarf elliptical galaxies are often found in the central 
regions of clusters of galaxies.
These observations suggest strongly
that the formation of most dwarf galaxies is linked to galaxy interactions. 
Therefore, now seems like the right time to investigate the formation
efficiency of such tidal dwarf galaxies.
Adopting the galaxy interaction scenario proposed by Silk \& Norman,
we find that if only a few dwarf galaxies are formed 
in each galaxy collision,
we are able to explain the observed morphology-density relations
for both dwarf and giant galaxies in the field,
groups of galaxies, and clusters of galaxies.
It seems worthwhile noting that the tidal dwarf formation may be 
coupled with the transformation from gas-rich disk galaxies
to early-type galaxies such as S0 and elliptical galaxies.
\end{abstract}


\begin{center}
To appear in ApJ, 543 (the November 1, 2000 issue)
\end{center}

\section{Introduction}              

Although dwarf galaxies are the most numerous systems in the nearby 
Universe (Ferguson \& Binggeli 1994; Binggeli, Sandage \& Tammann 1988;
Mateo 1998),
it seems unclear how they are related in origin to typical
giant galaxies\footnote{Contrasting to the naming of dwarf galaxies,
we call typical (i.e., ordinary-sized) galaxies ^^ ^^ giant galaxies"
in this article}.
Dwarf galaxies could be formed through the same physical process
formation of giant galaxies; i.e., gravitational collapse of 
protogalactic gas clouds (Dekel \& Silk 1986; White \& Frenk 1991;
Frenk et al. 1996; Kauffmann, Nusser \& Steinmetz 1997).
However, it is known that dwarf elliptical galaxies (dEs) apparently 
belong to the different class from giant ellipticals (Es)
in the fundamental plane (Kormendy 1985),
suggesting that the formation and/or evolution processes of dwarfs 
may not be the same as those of giants. 

It has been argued from an observational view point that
dwarf galaxies may be formed by galaxy collisions because 
there appears morphological evidence for dwarf galaxies in 
tidal tails of interacting galaxies (Zwicky 1956; Schweizer 1978);
i.e., gas-rich dwarf irregular galaxies can be made out of stellar and
gaseous material pulled out by tidal forces from the disks of colliding
parent galaxies into the intergalactic space. 
This possibility has been recently reinforced by a number of pieces of 
observational evidence
(Schweizer 1982; Bergvall \& Johansson 1985; Mirabel, Lutz \& Maza 1991;
Mirabel, Dottori \&  Lutz 1992; Duc \& Mirabel 1994,1998;
Yoshida, Taniguchi \& Murayama 1994; Braine et al. 2000;
Weilbacher et al. 2000; Hunter, Hunsberger, \& Roye 2000). 
Such formation of tidal dwarf galaxies has also
been demonstrated by numerical simulations
of merging/interacting galaxies 
(Barnes \& Hernquist 1992; Elmegreen, Kaufman \& Thomasson 1993).
Therefore, the tidal formation seems to be one of important formation
mechanisms of dwarf galaxies; note that these tidal dwarf galaxies 
(TDGs) are basically dwarf irregular galaxies when they are born.

The most populous type of dwarf galaxies is dEs
(Ferguson \& Binggeli 1994).
It is known that dEs generally trace 
the spatial distribution of giant galaxies in clusters of galaxies
and they tend to be close companions to giant galaxies
(Binggeli, Tarenghi \& Sandage 1990).
It should be also noted that the relative frequency of both dEs and giant
early-type galaxies is a monotonically increasing function of
the richness of groups and clusters
(Ferguson \& Binggeli 1994),
and this trend is found to be continued in less rich environments
such as the field (Vader \& Sandage 1991).
These observational results suggest that the formation of
dEs is strongly related to that of early-type galaxies
(E and S0 galaxies).
If the formation of early-type galaxies is related to 
galaxy interactions and/or merger events as suggested by 
Silk \& Norman (1981, hereafter SN81), a significant part of dEs 
may be fossils of TDGs which were formed through past tidal interactions.
Therefore, it seems important to
investigate a possibility that most early-type dwarf galaxies are also 
made by galaxy interactions.

%----------------------------------------------------------------------------

\section{Tidal Formation of Dwarf Galaxies}

\subsection{Model}

Let us suppose that galaxy interactions and/or merger events act 
as the dominant formation mechanism of dwarf galaxies in any environments.
In order to model this scenario, we adopt the galaxy interaction scheme 
proposed by SN81 because their scheme can be responsible for the observed 
morphology-density relation found by Dressler (1980). 
In this scenario, it is presumed that galaxy interactions occur
during the course of the hierarchical structure formation 
in the universe. 
In order to account for the tidal formation
of dwarf galaxies, we assume that some dwarf galaxies are made 
in each interaction.
This modified interaction scheme is shown in Table 1. Here, following SN81, 
we take account of only interactions between disk galaxies 
[spiral (Sp) and S0 galaxies].

Based on the above assumption, we obtain a set of kinetic equations for 
morphological type evolutions as a consequence of galaxy interactions 
in the following form:

\begin{eqnarray}
\label{one}
   & & \frac{1}{\gamma} \frac{dn_{\rm Sp}}{dt}
     = - 2 n_{\rm Sp}^2 - n_{\rm S0} n_{\rm Sp}, \\[10pt]
\label{two}
   & & \frac{1}{\gamma} \frac{dn_{\rm S0}}{dt}
     = n_{\rm Sp}^2 + ( 1 - 2 a ) n_{\rm Sp} n_{\rm S0}
                    - 2 b n_{\rm S0}^2, \\[10pt]
\label{thr}
   & & \frac{1}{\gamma} \frac{dn_{\rm E}}{dt}
     = b n_{\rm S0}^2 + a n_{\rm S0} n_{\rm Sp}, \\[10pt]
\label{fou}
   & & \frac{1}{\gamma} \frac{dn_{\rm dE}}{dt}
     = k_1 n_{\rm Sp}^2 + [ k_2 a + k_3 ( 1 - a ) ] n_{\rm S0} n_{\rm Sp} 
       + [ k_4 b + k_5 ( 1 - b ) ] n_{\rm S0}^2,
\end{eqnarray}
where $n_{\rm Sp}$, $n_{\rm S0}$,  $n_{\rm E}$, and $n_{\rm dE}$ are 
the number densities of spirals, S0s, ellipticals, and dwarfs, respectively,
$\gamma$ is the mean collision rate, and $k_i$ ($i = 1 - 5$) 
are the number of dwarfs formed by one collision in each case. 
Note that the first three equations are the same as those in SN81. 

In order to solve these equations, following SN81, we introduce the variable 
$x$ $\equiv$ $n_{\rm Sp}/n_{\rm S0}$ which 
decreases monotonically with increasing galaxy density.
Then we obtain the implicit 
solutions to equations (\ref{one}), (\ref{two}), (\ref{thr}), and (\ref{fou}) 
as below.

\begin{eqnarray}
\label{fiv}
   & & \frac{n_{\rm Sp}}{n_{0}} =
\Biggl[  \frac{x^2}{x^2 + ( 3 - 2 a ) x + 1 - 2 b} \Biggr] 
^{1/2 ( 1 - 2 b )} \nonumber \\[5pt]
   & & \quad \quad 
       \times {\rm exp} \Biggl[ \frac{1 + 2 a - 8 b}{( 1- 2 b ) \Delta}
       \Biggl({\rm tan}^{-1} \Biggl( \frac{2 x + 3 - 2 a}{\Delta} \Biggr)
                         - \frac{\pi}{2} \Biggr) \Biggr], \\[10pt]
\label{six}
   & & n_{\rm S0} = n_{\rm Sp} x^{-1}, \\[10pt]
\label{sev}
   & & n_{\rm E} = \int_x^{\infty}
           \frac{( b + a x ) n_{\rm Sp} dx}
                {x^2 [ x^2 + ( 3 - 2 a ) x + 1 - 2 b ]}, \\[10pt]
\label{eig}
   & & n_{\rm dE} = \int_x^{\infty}
           \frac{[ k_1 x^2 + { k_2 a + k_3 ( 1 - a ) } x
                  + k_4 b + k_5 ( 1 - b ) ] n_{\rm Sp} dx}
                {x^2 [ x^2 + ( 3 - 2 a ) x + 1 - 2 b ]},
\end{eqnarray}
where 
\begin{equation}
   \Delta^2 = 4 ( 1 - 2 b ) - ( 3 - 2 a )^2. 
\label{nin}
\end{equation}
Since $x$ decreases monotonically with increasing number density of 
giant galaxies, we obtain the following constraints on $a$ and $b$;
\begin{equation}
   \quad \quad 
   0 \leq b \leq \frac{1}{2}  -  \frac{1}{2}  (  \frac{3}{2} - a )^2,
   {\rm ~and~} \quad \frac{1}{2} \leq a \leq 1 .
\label{ten}
\end{equation}
We assume that all galaxies are initially spiral galaxies and their 
initial number density is $n_0$.
For simplicity, we suppose that the number of dwarfs formed by each 
collision case are identical in any cases; i.e.,  
$k_1=k_2=k_3=k_4=k_5=k$. Since our analysis is made from a statistical
point of view, we allow non-integer values of $k$ in later analysis. 
Then, we have only three parameters $a$, $b$, and $k$ in our model.
Our goal is to find a plausible solution responsible for observed number
densities of various types of galaxies as a function of galaxy number density.

It is uncertain whether or not interactions of gas poor S0s
produce similar dwarfs to spiral mergers. It seems likely that
stars liberated tidally from S0s may remain as a sparse stellar system
like low-surface-brightness (LSB) dwarfs.
However, since there is no such numerical simulation, we have no
firm answer to this question. Therefore, in our analysis,
we treat any kinds of dwarf galaxies as one population of 
dwarf galaxies. This treatment is enough when we compare our 
results with observations because no detailed
classification is made in the observational data given in 
section 2.2. However, this treatment will be refined in future
theoretical and observational studies. 

\subsection{Observational data}

We use the following two data sets. 1) Jerjen et al.'s (1992) sample
(hereafter the JTB sample):
They investigated galaxy populations brighter than $M=-15$
in Virgo cluster, Coma cluster, and 
some groups of galaxies using data compiled from the literature. 
Five morphological types (E, S0, Sp, Irr, and dE) are taken into account
in their analysis. In later analysis, irregular galaxies are included in the 
spiral sample.
2) Ferguson \& Sandage's (1991) sample (hereafter the FS sample):
They investigated galaxy populations in seven nearby groups of galaxies. 
Taking account of the completeness of their survey, we use the data of galaxies
with $M_{B_T} \leq -15.5$. 
Although the galaxies are classified morphologically  into nine types,
we classify dE, dE,N (nucleated dE), and dS0 as dE and Sa-Scd and Sd-Im as Sp
in our analysis. In Table 2, we give a summary of the two samples.
Since the limiting magnitudes are different between the two samples, 
we treat them as independent samples.

\subsection{Comparison of Model Results with the Observations} 

Here we solve the kinetic equations (1), (2), (3), and (4) numerically.
In these equations, there are the three free parameters;
$a$, $b$, and $k$. Taking the constraints in eq. (10), we vary $a$ from 0.5 to 1
with a step of 0.01 and $b$ from 0 to 0.37 with the same step\footnote{SN81
adopted $a=1$ and $b=0$ for simplicity.}.
Since it seems unlikely that more than several tidal dwarfs are formed in one
interaction, we vary $k$ from 0 to 5 with a step of 0.1.
In order to find the best model, we minimize the 
the sum of square of deviations 
between the prediction and the observation
weighted by the predicted number of galaxies;

\begin{equation}
\delta = \sum_{i=1}^{4} \sum_{j=1}^{N}
\frac{(n_{\rm pred} - n_{\rm obs})_{i,j}^2}{n_{\rm pred}}
         = \sum_{i=1}^{4} \sum_{j=1}^{N}
\frac{\chi^2}{n},
\label{twel}
\end{equation}
where the sum of $i$ is done for 4 morphological types
(i.e., E, S0, Sp, and dE) and
that of $j$ is done for the data of groups and clusters,
$N$ is the total number of galaxies for each type,
and $n_{\rm pred}$ and $n_{\rm obs}$ are the numbers of 
predicted and observed galaxies.
The best estimate of $k$ is determined by the observed dwarf fractions
and weakly depend on the fractions of E, S0, and Sp. On the other hand,
the best estimates of $a$ and $b$ are determined 
by the giant fraction in each type. 

For the JTB sample, we obtain the best parameter set of $a=1$, $b=0$ and $k=2.0$.
However, for the FS sample, although we cannot obtain a definitive model, 
we find that 
the model with the least value of $\delta$ is the one with $a=1$, $b=0.37$,
and $k=0.8$. The reliability of this result seems worse than 
that based on the JTB sample because the FS sample covers a smaller range in $x$.
These model results are compared with the observations for the JTB and FS
samples in Fig. 1. We show the observed 
fractional abundances of the individual morphology types together with
the model predictions by curves.
For the JTB sample, our model appears consistent with 
the observed increase of both the early-type giants and dEs.
Ferguson \& Sandage (1991) reported that the number of dEs increases 
with the richness of groups and clusters. 
Together with this, our results suggest that the 
dwarf formation efficiencies ($k$) may be more affected by their environments 
rather than by the {\it local} value of $x$ because 
$x$ does not always correspond to 
the richness of groups and clusters. 
For example, the Virgo cluster is the system with both the largest number of 
galaxies and the lowest $x$ while N1400 group is a relatively poor group with the 
largest $x$.

%---------------------------------------------------------------------------

\section{Discussion}

We have found that our model can be responsible for the observed numbers of
dEs in the various environs from poor groups of galaxies to usual rich clusters
of galaxies. The formation rate of TDGs is estimated to be $\sim$ 1 -- 2 in each
galaxy interaction. It is interesting to compare this value with the actual
observed numbers of TDGs. In Table 3, we give a summary of the observed 
number of TDGs, $N_{\rm TDG}$, in the seven interacting/merging galaxies.
Although about 10 TDGs are detected in NGC 5291 (Duc \& Mirabel 1998) 
and the Superantennae (Mirabel, Lutz \& Maza 1991)
a few TDGs are typically observed in the remaining five galaxies. 
Although the three TDGs in ESO 148$-$IG02 (Bergvall \& Johansson 1985) 
are relatively bright,
$M_B \sim -17$ -- $-18$, most the remaining TDGs are fainter than 
$M_B \sim -15$ (e.g., Yoshida, Taniguchi \& Murayama 1994). 
Since we use the data
of galaxies brighter than $M_B \sim -15$ in the present study, 
our result ($k \sim$ 1 -- 2) appears consistent with the observations.

Barnes \& Hernquist (1992) have shown that a dozen TDGs can be formed
in the tidal tails of a merger between two gas-rich disk galaxies.
The most massive objects among them reach several $\times 10^8 M_\odot$
which are expected to be brighter than $M_B \sim -15$. However, it is noted
that the majority of the TDGs in their simulations are fainter than 
$M_B \sim -15$. Therefore, our results also appear consistent with
the numerical simulations. 

The galaxy interaction scheme of SN81 adopted in this study is basically 
addressed to the early collapse phase of clusters of galaxies at high redshift.
Therefore, even if a numerous TDGs could be formed in each interaction
between protodisk galaxies, a significant part of them would either merge to form
more massive objects or return to some giant galaxies
during the course of dynamical evolution of clusters of galaxies.
Therefore, our result (i.e., $k \sim$ 1 -- 2) should be regarded as
the number of TDGs which have been surviving for $\sim 10^{10}$ years.
Even if a large number of TDGs were originally formed in early galaxy
interactions, some of them could merge to form more massive objects or
return either to the parent or to some neighboring galaxies.

Finally it is interesting to note that the tidal formation of dwarf 
galaxies can be responsible for the dichotomy of dEs; i.e., nucleated
or non-nucleated dEs. As demonstrated by the numerical simulations,
the most massive objects tend to contain
a lot of gas with respect to the stellar content.
Even if the gas is dominated by HI gas, the molecular gas formation could
proceed from the HI dominated gas as demonstrated by the recent CO
observations of the TDGs. Therefore, it seems very likely
that the intense star formation
could occur in the center of such tidal dwarfs.
The important point is that the resultant star cluster is expected
to be dynamically decoupled from the remaining stellar system. 
Then it is expected that gas-rich tidal dwarfs could evolve to nucleated dEs.
On the other hand, less-massive TDGs tend to have little gas and thus 
they evolve to non-nucleated dEs as suggested by Barnes \& Hernquist (1992).
In conclusion, the tidal formation scenario presented here has a number of
merits to explain the observational properties of dwarf galaxies.

\vspace {0.5cm}

We would like to thank an anonymous referee for useful suggestions
and comments.
This work was financially supported in part by
the Ministry of Education, Science, and Culture
(Nos. 10044052, and 10304013).

%----------------------------------------------------------------------
%           References
%----------------------------------------------------------------------

%----------------------------------------------------------------------

%%%%%%%%%%%%%%%%%%%%%%%%%%%%%%%%%%%%%%%%%%%%%%%%%%%%%%%%%%%%%%%%%%%%%%%%
%                         Table (1)
%%%%%%%%%%%%%%%%%%%%%%%%%%%%%%%%%%%%%%%%%%%%%%%%%%%%%%%%%%%%%%%%%%%%%%%%

%%%\begin{table}[t]
%%%\caption{merger Scheme} 
\begin{center}
{\bf Table 1.}
Merger Scheme \\
\vspace{0.5cm}
\begin{tabular}{ lccl } \hline \hline \\
1. & Sp + Sp & $\longrightarrow$           & S0 + $k_1$ dE       \\
2. & Sp + S0 & $\longrightarrow^{\rm (a)}$ & E + $k_2$ dE       \\
   &         & $\searrow^{\rm (1-a)}$      & S0 + S0 + $k_3$ dE \\
3. & S0 + S0 & $\longrightarrow^{\rm (b)}$ & E + $k_4$ dE       \\
   &         & $\searrow^{\rm (1-b)}$      & S0 + S0 + $k_5$ dE \\
\hline \hline \\
\end{tabular}
\end{center}
%%%\end{table}%
\vspace{0.5cm}
In this scheme, a merger between two spiral galaxies evolves not into an
an elliptical galaxy but into an S0 one. The reason for this is as follows.
It is widely accepted that elliptical-like products are formed 
by dissipationless collapse.
Mergers between gas-rich spiral galaxies can achieve a similar
physical condition in their final phase.
However, if the star formation timescale is longer significantly than the 
dynamical timescale, 
the remained gas will settle to a disk and then the end product will 
not become an elliptical-like galaxy. 
This is confirmed by analytical and numerical methods. 

%%%%%%%%%%%%%%%%%%%%%%%%%%%%%%%%%%%%%%%%%%%%%%%%%%%%%%%%%%%%%%%%%%%%%%%%
%                         Table (2)
%%%%%%%%%%%%%%%%%%%%%%%%%%%%%%%%%%%%%%%%%%%%%%%%%%%%%%%%%%%%%%%%%%%%%%%%
%%%\begin{table}[t]
%%%\caption{Observational Morphological Populations}
\begin{center}
{\bf Table 2.}
Observational Morphological Populations \\
\vspace{0.5cm}
\begin{tabular}{ ccccc } \hline
       &   E   &   S0   &    Sp  &  dE   \\ \hline\hline
\multicolumn{5}{c}{JTB sample} \\ \hline
groups &   3   &  6   & 88   &  3  \\ 
Virgo  &   6   &  9   & 40   & 45  \\
Coma   &  10   & 12.5 &  2.5 & 75  \\ \hline
\multicolumn{5}{c}{FS sample} \\ \hline
Leo    &  20.8 & 15.0 & 47.5 &  8.3 \\
Dorado &  14.1 & 23.3 & 50.3 &  9.2 \\
N1400  &  18.6 & 27.5 & 15.0 & 39.0 \\
N5044  &  21.6 & 15.5 & 27.2 & 31.5 \\
Fornax &  15.4 & 15.5 & 33.5 & 31.9 \\
Antlia &  11.7 & 14.6 & 30.4 & 41.0 \\
Virgo  &   7.4 & 10.8 & 37.9 & 38.7 \\ \hline
\end{tabular}
\end{center}
%%%\end{table}%
\vspace{1.0cm}

%%%%%%%%%%%%%%%%%%%%%%%%%%%%%%%%%%%%%%%%%%%%%%%%%%%%%%%%%%%%%%%%%%%%%%%%
%                         Table (3)
%%%%%%%%%%%%%%%%%%%%%%%%%%%%%%%%%%%%%%%%%%%%%%%%%%%%%%%%%%%%%%%%%%%%%%%%
%%%\begin{table}[t]
\begin{center}
{\bf Table 3.}
Observed numbers of TDGs  \\
\vspace{0.5cm}
\begin{tabular}{ llcl } \hline \hline \\
No. &  Galaxy  &   $N_{\rm TDG}$  &   Reference  \\ \hline
1 & NGC 7252 & 2 & Schweizer (1982) \\
2 & ESO 148$-$IG02 & 3 & Bergvall \& Johansson (1985)  \\
3 & The Antennae$^a$ & 1 & Mirabel et al. (1992)      \\
4 & The Superantennae$^b$ & 9 & Mirabel et al. (1991) \\
5 & Arp 105 & 2 & Duc \& Mirabel (1994) \\
6 & NGC 2782 & 1 & Yoshida et al. (1994) \\
7 & NGC 5291 & 11 & Duc \& Mirabel (1998) \\
\hline \hline \\
\end{tabular}
\end{center}

$^a$ NGC 4038 + NGC 4039

$^b$ IRAS 19254$-$7245 

%------------------------------------------------------------------------------

\begin{figure}
\epsfysize=19.0cm \epsfbox{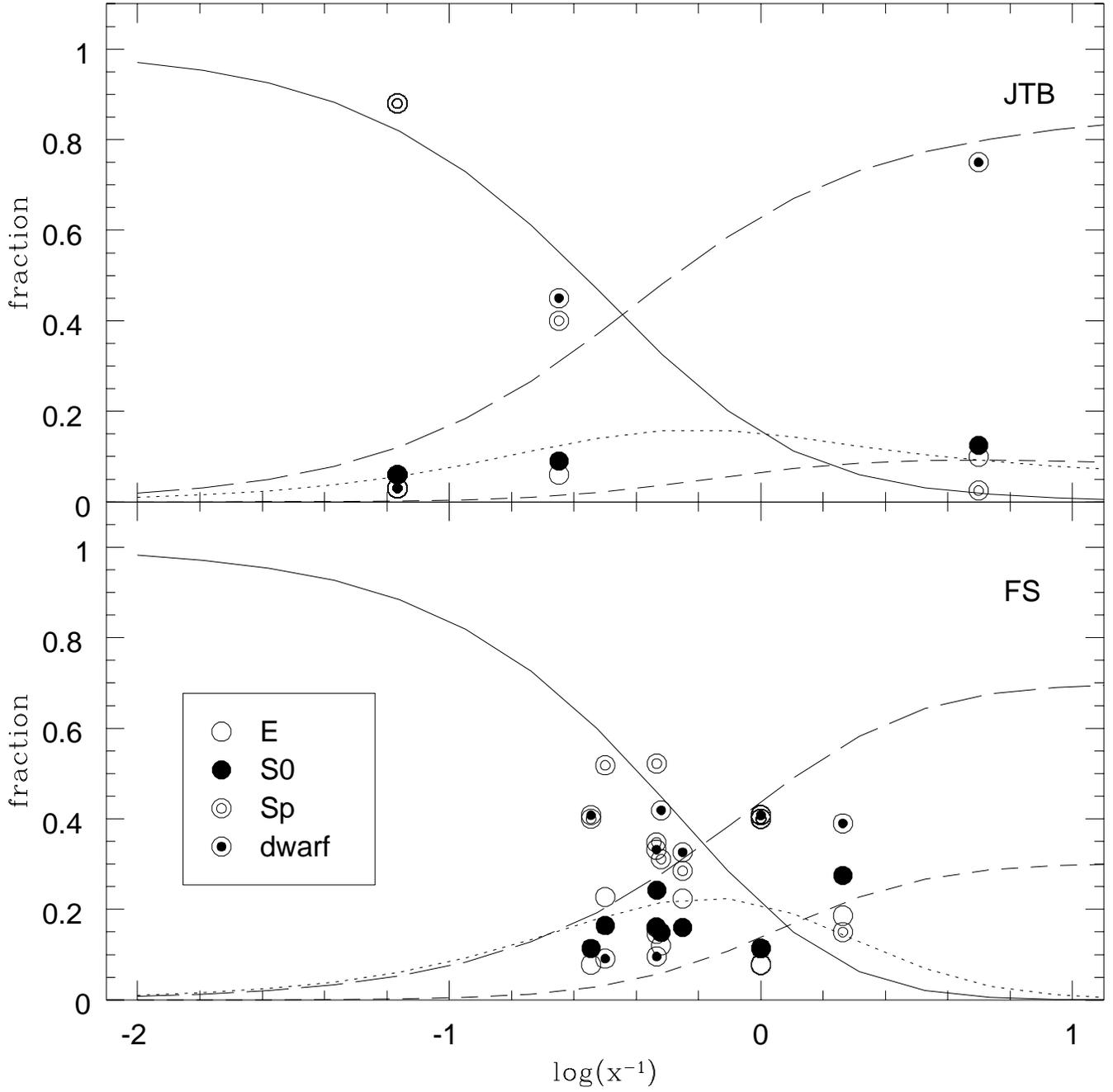}
\caption[]{
The best model for the JTB sample (upper panel) and that
for the FS sample (lower panel).
Fractional abundances of E (open circles), Sp (double open circles),
S0 (filled circle), and dEs (filled-in open circle) are shown as
a function of $x$.
\label{fig1}
}
\end{figure}

%------------------------------------------------------------------------------

\end{document}